\documentstyle[12pt,aasms]{article}
\begin{document}
\title{THE TINSLEY DIAGRAM REVISITED}

\author{Stephen Leonard and Kayll Lake\altaffilmark{1}}
\affil{Department of Physics, Queen's University at Kingston, Kingston,\\
Ontario, Canada K7L 3N6}
\altaffiltext{1}{E-mail:  lake@astro.queensu.ca}

\begin{abstract}

Motivated by the recent determinations of the Hubble constant $(H_0)$
from observations of Cepheid variables in NGC4571 and M100, we plot the
Tinsley diagram with level curves of the cosmological constant $(\Lambda
)$. Based on current estimates of the absolute ages of globular clusters
we conclude that $\Lambda > 0$
and, irrespective of the background spatial curvature, the universe will
not recollapse. These
conclusions hold for both relativistic and Newtonian models and are
{\it independent} of the density parameter.
\end{abstract}


\section{Introduction}
Tinsley's modification of the Robertson diagram (\cite{tin67}), the plot
of $H_0t_0 ~{\rm vs.}  \log (\Omega_0/2)$, where $t_0$ is the age of the
universe
and $\Omega_0$ the current density parameter, allows us to limit the
cosmological
parameters simply because $t_0$ must be greater than the oldest
observed objects. (See, for example \cite{rin77} for an introduction.) In
view of the recent determinations of $H_0$ from the direct observations
of Cepheid variables in NGC4571 (\cite{pie94}) and M100 (\cite{fre94})
in the Virgo cluster it would seem an appropriate time to reexamine the
Tinsley diagram. In this {\it Letter} we construct the diagram based on a
relativistic model consisting of a non-interacting mixture of dust and
blackbody radiation. (This is known to be an excellent approximation to a
baryon-conserving equilibrium mixture of blackbody photons and a neutral
relativistic Maxwell-Boltzmann gas for an initial $f(\equiv n\gamma
/n_b$,  photon to baryon
number density ratio) $>$ 100 which is most certainly the case.  Other
(non-baryonic) species are handled analogously.)

\section{Construction of the Diagram}
\subsection{Relativistic model}
Friedmann's equation for the evolution of a separately conserved
distribution of dust and blackbody radiation gives rise to the relation
\begin{equation}
ht_0\propto \int^1_0 {wdw\over\sqrt{\lambda w^4 + (1 - \lambda)w^2 -
(\Omega_0 + \tau_0)w^2 + \Omega_0w + \tau_0}}
\end{equation}
where $h {,}~ \lambda {,}~ \Omega_0$, and $\tau_0$ are
dimensionless parameters. We measure $t_0$  in units of 10 Gyr. The
proportionality constant turns out to be 1/1.02. (In what follows we
display three significant figures). The parameters are defined by
\begin{equation}
h = H_0/100
\end{equation}
where $H_0$  is in km s$^{-1}$ Mpc$^{-1}$, and
\begin{equation}
\lambda = 2.85 ~10^{55}~\Lambda /h^2
\end{equation}
where $\Lambda$ is in cm$^{-2}$.  $\tau_0$ is defined by
\begin{equation}
\tau_0 = 32 \pi Gc \sigma T^4_0/3H^2_0
\end{equation}
where $\sigma$ is the Stephan-Boltzmann constant, $G$ is Newton's constant,
$T_0$ is the current background temperature, and $c$ is the speed of light.
We find
\begin{equation}
\tau_0 = 2.38 ~10^{-5}/h^2.
\end{equation}
$\Omega_0$ is defined by
\begin{equation}
\Omega_0 = 320 \zeta (3) GT^3_0(m_e + m_p) \sigma/\pi^3 fckH^2_0
\end{equation}
where $\zeta$ is Riemann's function, $k$ is Boltzmann's constant, and $m_e$ and
$m_p$ give the
electron and proton masses. We find
\begin{equation}
\Omega_0 = 3.56 ~10^7/f h^2.
\end{equation}
The spatial curvature of the background is determined by $\epsilon$ where
\begin{equation}
{\rm sign}(\epsilon )  = {\rm sign }(\Omega_0 + \tau_0 + \lambda - 1).
\end{equation}
We have relegated the deceleration parameter $q_0$ to a supplementary
role. It is determined by the relation
\begin{equation}
q_0 = \Omega_0/2 + \tau_0 - \lambda .
\end{equation}
For other number-conserving massive particle species which can be
treated as dust, we replace $f$ by $n_\gamma /n$ and $m_e + m_p$ by
$m$ with $n\equiv \Sigma_sn_s$ and $m \equiv \Sigma_sm_s$ where we
sum over all particle species.  Whereas the inclusion of non-baryonic
species changes the interpretation of $f$ and the numerical coefficient in
relation (7), the actual value of $\Omega_0$ does not enter the arguments
which follow.
\subsection{Newtonian model}

It turns out that $\tau_0$ makes no significant contribution to the
integral (1) for log$_{10} ~ (\Omega_0/2) > -2.5$ (the range we consider).
With
$\tau_0 = 0$ the integral (1) can be obtained directly from a Newtonian
model with the Poisson equation generalized to read as
\begin{equation}
\nabla^2 \Phi + c^2\Lambda = 4\pi G\rho .
\end{equation}
The integral (1) follows from the integral of the energy along with the
equation of continuity. In this model
\begin{equation}
\Omega_0 = 5.32 ~10^{28}\rho_0/h^2
\end{equation}
where $\rho_0$ is the Newtonian mass density in g cm$^{-3}$ and includes
all species contributing to the "dust".

\section{The Diagram}
\subsection{Properties}
Numerical integration of the integral (1) results in Figure 1. The ordinate
$y$ is 1.02$ht_0$ and the abscissa log$_{10}(\Omega_0/2)$ where
$\Omega_0$ is interpreted by relation (7) or (11). The range in $x$ shown
covers all possibilities as $x < -2.3$ violates the primordial production of
deuterium and $^3$He. Note that $\Omega_0 = 1$ corresponds to $x = -
0.30$. Level curves of constant $\lambda$ are shown as is the locus
$\epsilon = 0$. It
is important to note that models which recollapse lie below $\Lambda =
0$ up to its intersection with $\epsilon = 0$ and below the locus
\begin{equation}
27\lambda\Omega^2_0 = 4(1 - \lambda - \Omega_0 - \tau_0)^3
\end{equation}
thereafter.  For the range shown the locus (12) lies close to $\Lambda =
0$.

\subsection{Limits}

Clearly $y > 1.02ht^\ast \equiv y^\ast$ where $t^\ast$  corresponds
to the oldest absolute age observed. Given $h$, this may be viewed as an
absolute bound on $\lambda$ and $\epsilon$ over the range of $x$ shown.
We choose to determine $t^\ast$ by the absolute ages of globular
clusters.

According to \cite{pie94}
\begin{equation}
h = 0.87 \pm 0.07 ,
\end{equation}
and according to \cite{fre94}
\begin{equation}
h = 0.80 \pm 0.17.
\end{equation}
\cite{cha94} (see also \cite{cha194}) gives
\begin{equation}
t^\ast = 1.55 \pm 0.4 ,
\end{equation}
whereas \cite{van91} gives
\begin{equation}
t^\ast = 1.65 \pm 0.2 .
\end{equation}
 From the observations (13) and (14) with the ages (15) and (16) we find
\begin{equation}
y^\ast = 1.36 \pm 0.36
\end{equation}
where we have weighted all opinions equally and quoted the standard
deviation associated with the sixteen extreme values. From Figure 1 and
the estimate (17) we obtain the following conclusions:

{\it i) The universe will not recollapse,}

and

{\it ii)}  $\Lambda > 0.$

It should be noted that conclusion i) holds even if $\epsilon  > 0.$

\section{Discussion}
It is common knowledge that a high value of $h$, given the ages of
globular clusters, would lead to $\Lambda > 0$. The purpose of this Letter
is to make the argument, which is independent of the value of $\Omega_0$,
clear. There
is a well known resistance to $\Lambda$ on both aesthetic and
philosophical grounds, the latter being encapsulated by the argument that
$\Lambda$ affects all matter but is affected by nothing (for example,
\cite{col94}). However, one could also argue, on the basis of the
 Weyl-Lovelock theorem (\cite{lov71}), that $\Lambda$ is an essential
element of the
underlying theory. This theorem states that in four dimensions the most
general symmetric tensor field $A_{ij} = 0$ which is divergence free and
constructed out of the metric tensor and its first two (partial)
derivatives is $A_{ij} = G_{ij} + \Lambda g_{ij}$. We are no more justified
in dismissing $\Lambda$ on the basis of its size ($\sim$  few $10^{-56}$
cm$^{-2}$ ) than we are in dismissing quantum mechanics because of the
value of Planck's constant.  However, if $\Lambda > 0$, as the empirical
evidence would suggest, a number of fundamental results in general relativity
(including the uniqueness theorems) will have to be reexamined.

We thank Brian Chaboyer for the estimate (15) and Dave Hanes for a
preprint of Freedman et al. 1994.

\begin{figure}
\caption{The Tinsley diagram plotted with level curves of constant
$\lambda$ . The curves shown correspond to $\lambda = -3{,}~ -2{,}~ -
1{,}~
0{,} ~1{,}~ 2{,}~ 3{,}~ 3.5{,}~ 4{,}$ and 5. The locus $\epsilon  = 0$ is also
shown.}
\end{figure}

     {\it Note added in proof.-} Accoring to X. Shi (private communication,
and preprint [1994]) a high helium abundance (e.g. $Y \approx 0.28$), as well
as helium diffusion and oxygen-enhancement in stellar models could lower $t^*$
to about $1.1 \pm 0.1$. With the observations (13) and (14) this would lower
$y^*$ to $0.92 \pm 0.18$.

\begin{thebibliography}{}
\bibitem[Chaboyer (1994)]{cha94}\reference Chaboyer, B. 1994, private
communication
\bibitem[Chaboyer  {\it et al.} (1994)]{cha194}\reference Chaboyer, B.
Demarque, P., Guenther, D. B.,
Pinsonneault, M. H., \& Pinsonneault, D. B. 1994,
preprint
\bibitem[Coles and Ellis (1994)]{col94}\reference Coles, P. \& Ellis, G.
1994, preprint
\bibitem[Freedman   {\it et al.}  (1994)]{fre94}\reference Freeman, W. L.
et al. 1994 Nature, 371, 757
\bibitem[Lovelock (1971)]{lov71}\reference Lovelock, D. 1971, J. Math.
Phys., 12, 498
\bibitem[Pierce  {\it et al.} (1994)]{pie94}\reference Pierce, M. J. {\it et
al.} 1994 Nature, 371, 385
\bibitem[Rindler  (1977)]{rin77}\reference Rindler, W. 1977, Essential
Relativity (New
York: Springer-Verlag)
\bibitem[Tinsley (1967)]{tin67}\reference Tinsley, B. M. 1967, Ph.D.
dissertation, University of Texas
\bibitem[VandenBergh (1991)]{van91}\reference VandenBergh, D. A. 1991,
in The Formation and
Evolution of Star Clusters, PASP Conf. Ser. 13
ed.K. Janes (San Francisco:Astr. Soc. Pacific), 183
\end{thebibliography}
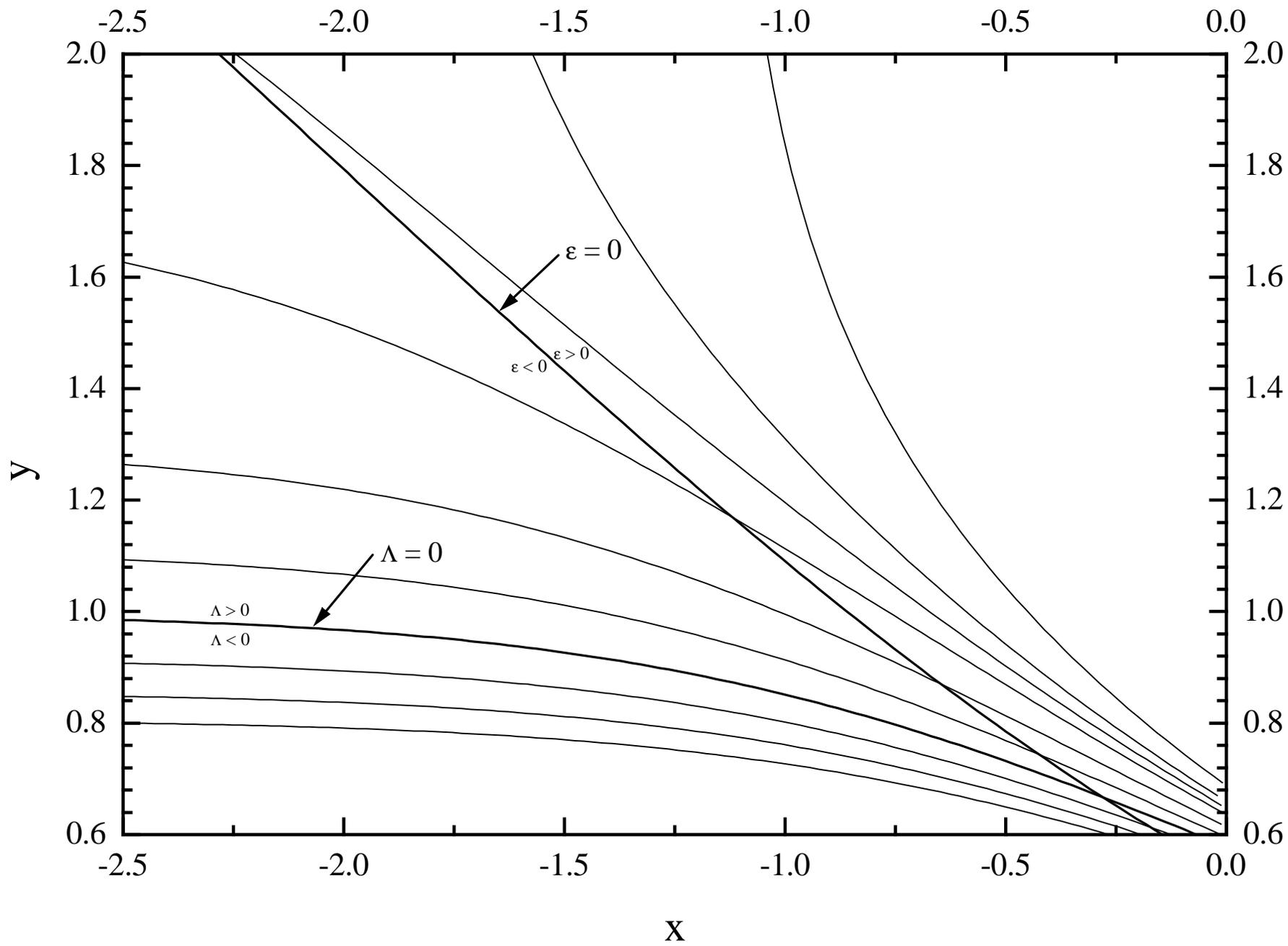
\end{document}